\documentclass[twocolumn,epjc3]{svjour3}
\usepackage{graphicx}
\usepackage{epsfig}
\usepackage{amsmath,amssymb}
\journalname{Eur. Phys. J. C}
\begin{document}
\title{Pauli-Zeldovich cancellation of the vacuum energy divergences, auxiliary fields and supersymmetry}
\author{Alexander~Yu.~Kamenshchik\thanksref{e1,addr1,addr2},
Alexei~A.~Starobinsky\thanksref{e2,addr2,addr3},
Alessandro~Tronconi\thanksref{e3,addr1},
Tereza~Vardanyan\thanksref{e4,addr1}
\and
Giovanni~Venturi\thanksref{e5,addr1}}
\thankstext{e1}{e-mail: kamenshchik@bo.infn.it}
\thankstext{e2}{e-mail: alstar@landau.ac.ru}
\thankstext{e3}{e-mail: tronconi@bo.infn.it}
\thankstext{e4}{e-mail: tereza.vardanyan@bo.infn.it}
\thankstext{e5}{e-mail: giovanni.venturi@bo.infn.it}
\institute{Dipartimento di Fisica e Astronomia, Universit\`a di Bologna and INFN,
via Irnerio 46, 40126 Bologna, Italy \label{addr1}
\and
L. D. Landau Institute for
Theoretical Physics, Moscow, 119334 Russia\label{addr2}
\and
National Research University Higher School of Economics, Moscow, 101000 Russia
\label{addr3}}

\date{Received: date / Accepted: date}
\maketitle

\begin{abstract}
We have considered the Pauli-Zeldovich mechanism for the cancellation of the ultraviolet divergences in vacuum energy. This mechanism arises  because bosons and fermions give  contributions of the opposite signs. In contrast with the preceding papers devoted to this topic wherein mainly free fields were studied, here we have taken their interactions into account to the lowest order of  perturbation theory. We have constructed some simple toy models having particles with spin $0$ and spin $1/2$, where masses of the particles are equal while the interactions can be quite non-trivial.
\end{abstract}

\section{Introduction}
Many years ago Pauli~\cite{Pauli} suggested that the vacuum (zero-point) energies of all existing fermions and bosons compensate each other. This possibility is based on the fact that
vacuum energy of fermions has a negative sign whereas that of bosons has a positive one. As is well known, such a cancellation indeed takes place in supersymmetric models (see e.g. \cite{supersymm}). Subsequently in a series of papers Zeldovich~\cite{zeld,zeld1} related 
vacuum energy to the cosmological constant. However rather than eliminating divergences through the boson-fermion cancellation, he suggested the Pauli-Villars regularization of all divergences by introducing a number of massive regulator fields. Covariant regularization of
all contributions then leads to finite values for both the energy density $\varepsilon$ and (negative) pressure $p$ corresponding to a cosmological constant, i.e. connected by the equation of state $p =-\varepsilon$. 

In our preceding paper \cite{we} we examined the conditions for the cancellation of the ultraviolet divergences of the vacuum energy to the leading order in $\hbar$, i.e. by considering free theories and neglecting interactions. Such conditions are reduced to some sum rules involving the masses of particles present in the model. We formulated these conditions not only for the Minkowski spacetime, but also for the de Sitter one. In the latter case, the radius of the de Sitter universe also enters into the mass sum rules. In paper \cite{we1} we applied such considerations to observed
particles of the Standard Model (SM) and also studied the finite part of vacuum
energy. This last contribution should be very small, 
so as to obtain a result compatible with the observed value of the cosmological constant
(almost zero with respect to SM particle masses). 
We showed \cite{we1} that it was impossible to construct a minimal extension of the
SM by finding a set of boson fields which, besides canceling ultraviolet
divergences, could compensate residual huge contribution of known
fermion and boson fields of the Standard Model to the finite part of the
vacuum energy density.

On the other hand, we found that addition of at least one massive fermion field
was sufficient for the existence of a suitable set of boson fields which would permit
such cancellations and obtained their allowed mass intervals. On examining
one of the simplest SM extensions satisfying the constraints, we found that the mass 
range of the lightest massive boson was compatible with the Higgs mass bounds which 
were known at the time of the publication of the paper 
\cite{we1}. As is well known, later the Higgs boson was discovered at the LHC \cite{Higgs,Higgs1}. For some time it appeared that there 
might exist an the observed diphoton excess at 750 \rm{GeV} \cite{diphoton}. 
This excess, had it been confirmed, could be interpreted as an indication for the existence 
of a new heavy elementary or composite particle with  a mass of the order of 750 \rm{GeV}. Later this phenomenon  disappeared, 
nonetheless inspiring in the meanwhile quite a few theoretical works.  
In particular, we also studied in our preprint how 
the presence of such a new particle could be included into our scheme of the cancellation of ultraviolet divergences of vacuum energy \cite{we-750}.  

In our preceding papers \cite{we,we1,we-750} we studied only free theories without interactions. It is also interesting to take interactions into account, at least to the lowest order of  perturbation theory. This is not easy, and in the present paper we shall concentrate on the construction of relatively simple toy models where 
the ''Pauli-Zeldovich cancellation'' of ultraviolet divergences still takes place.  

We wish to emphasize that the approach employed in the present paper represents a 
whole direction in quantum field theory which goes well beyond effective low energy field theory and, although based on some hypothesis, has not been proven to be wrong.  We also wish  to mention the paper by Ossola and Sirlin \cite{Sirlin}, where contributions of fundamental particles to the vacuum energy density were discussed with a special attention to relations between different regularization schemes and to the appearance of power divergences in different contexts. Other related approaches are presented in Refs. \cite{Nambu2,Veltman,Dudas}.

In the recent paper \cite{Visser},  it was noticed that under certain circumstances (in particular, but not limited to finite QFTs), the Pauli cancellation mechanism would survive the introduction of particle interactions. It was pointed out there that for the mass sum rules to be valid at different mass scales, it is necessary to impose some relations on mass runnings with energy. Thus, the corresponding relations between anomalous mass dimensions were formulated \cite{Visser}. However, concrete examples were not constructed. 

In the present paper we discuss some relatively simple examples of models where the Pauli-Zeldovich cancellation takes place to the first order of  perturbation theory. Being inspired by the famous supersymmetric Wess-Zumino model \cite{WZ}, we consider models with spinor, scalar and pseudoscalar fields only. We hope to treat  vector (gauge) fields in future works. The models which we discuss are not supersymmetric, but they have  one important feature which makes them akin to supersymmetric models: the number of the fermion and boson degrees of freedom in them is the same. That implies an unexpected feature: the necessity to take so called auxiliary fields into account. Such fields are necessary in the supersymmetric models, because they allow one  to formulate supersymmetry transformations in a coherent way.

But their role is even more ubiquitous. To conserve supersymmetry, it is necessary to have the balance between fermion and boson degrees of freedom not only on shell, but also off shell. However, the number of degrees of freedom of a spinor field doubles when it is off shell. For example, a Majorana spinor has two complex components, i.e. four degrees of freedom off shell. When we require the satisfaction of the first-order Dirac equation, the number of degrees of freedom becomes equal to two. Thus, for example, in the Wess-Zumino \cite{WZ} model one has two fermion degrees of freedom of the Majorana spinor and two boson degrees of freedom associated with the scalar and pseudoscalar fields. Off shell the number of fermion degrees of freedom becomes equal to four, while the role of two additional boson fields is played by two auxiliary fields which become in a sense independent off shell. If we consider non-supersymmetric models with the Pauli-Zeldovich  mechanism of  cancellation of ultraviolet divergences for vacuum energy in the presence of interactions, then the number of the boson and fermion degrees of freedom should be equal not only on shell, but also off shell. This means that we should introduce auxiliary fields. Further, when we consider a model with  interactions, we should not only take into account running of masses of the fields, but also consider cancellations of contributions coming from the potential terms in the Lagrangians. It is there that the introduction of the auxiliary fields becomes very convenient. Fortunately, we shall see that, at least in the considered class of spinor-scalar models, the introduction of auxiliary fields is equivalent to a simple rule for the calculation of some contribution to the scalar fields self-interaction. Here we can add that, in principle, one can perform all calculations and show that in the formalism where auxiliary fields are excluded, vacuum energy in the supersymmetric models is equal to zero. However, in this case there are no separate cancellations of the potential energy and of the kinetic energy between  bosons and fermions. Thus, verification of the analogous cancellation in non-supersymmetric models becomes more complicated. Hence, it is better to implement rather simple rules, equivalent to the explicit introduction of auxiliary fields, which will be used in the present paper.    

Here we present a model consisting of  a Majorana fermion and two scalar fields with the same mass and with different kinds of interactions, and we show that for such a model, one can find a family of  coupling constants such that  the Pauli-Zeldovich mechanism for the cancellation still works. Then we find an analogous family of models with a Majorana fermion, a scalar field and a pseudoscalar field. Obviously, the Wess-Zumino model belongs to this family. We also discuss briefly models where particles with different masses are present. 

The structure of the paper is as follows: in the second section we briefly discuss  the mass sum rules for theories without interactions; in the third section we formulate rules for the conservation of the mass sum rules when interactions are switched on. In Sec. 4 we discuss the vacuum expectation values of the potential terms and the role of auxiliary fields. In  Sec. 5 we present a model with one Majorana field and two scalar fields. In the sixth section we consider a model with one Majorana field, one scalar field and one pseudoscalar field. Sec. 7 is devoted to the discussion of models with non-degenerate masses, the last section contains some concluding remarks. 

\section{Vacuum energy and the balance between the fermion and boson fields}

One knows that vacuum energy of the harmonic oscillator is equal to $\frac{\hbar\omega}{2}$. If one has  a massive  field with  mass $m$, then $\omega =\sqrt{k^2c^2 + m^2c^4}$, where $k$ is the wave number.  In the following we shall set $\hbar=1$ and $c=1$. The energy density of vacuum energy of a scalar field treated as free oscillators with all possible momenta is given by the divergent integral \cite{zeld}:
\begin{equation}
\varepsilon=\frac12\int d^3k\sqrt{k^2+m^2} = 2\pi\int_0^{\infty}dk k^2 \sqrt{k^2+m^2}. 
\label{en}
 \end{equation}
We can regularize this integral by introducing a cutoff $\Lambda$. In this case
\begin{eqnarray}
&&\varepsilon=2\pi\int_0^{\Lambda}dk k^2 \sqrt{k^2+m^2}\nonumber \\
&&=2\pi m^4\left[\frac{\Lambda}{8m}\left(\frac{2\Lambda^2+1}{m^2}\right)\sqrt{\frac{\Lambda^2}{m^2}+1}\right.\nonumber \\
&&\left.-\frac18\ln\left(\frac{\Lambda}{m}+\sqrt{\frac{\Lambda^2}{m^2}+1}\right)\right].
\label{en1} 
 \end{eqnarray}
On expanding this expression with respect to the small parameter $\frac{m}{\Lambda}$, one obtains
\begin{equation}
\varepsilon=\frac{\pi}{2}\Lambda^4+\frac{\pi}{2}\Lambda^2m^2+\frac{\pi}{16}m^4(1-4\ln2)-\frac{\pi}{4}m^4\ln\frac{\Lambda}{m}+o\left(\frac{m}{\Lambda}\right).
\label{en2}
\end{equation}
The contribution of one fermion degree of freedom coincides with that of Eq. (\ref{en})  with the opposite sign. 
It now follows from Eq. (\ref{en2}) that to cancel the quartic ultraviolet divergences proportional to $\Lambda^4$, one has to have equal numbers of boson and fermion degrees of freedom:
\begin{equation}
N_B=N_F. 
\label{en0}
\end{equation}
The conditions for the cancellation of quadratic and
logarithmic divergences  are
\begin{equation}
\sum m_S^2 +3 \sum m_V^2 = 2 \sum m_F^2
\label{quad}
\end{equation}
and
\begin{equation}
\sum m_S^4 +3 \sum m_V^4 = 2 \sum m_F^4\,,
\label{log}
\end{equation}
respectively.
Here the subscripts $S$, $V$ and $F$ denote scalar, massive vector and massive
spinor Majorana fields respectively (for Dirac fields it is sufficient to put a $4$
instead of $2$ on the right-hand sides of Eqs. (\ref{quad}) and (\ref{log})).
For the case in which the conditions (\ref{en0}), (\ref{quad}) and (\ref{log}) are satisfied, the remaining finite part of the vacuum energy density is  
equal to
\begin{equation}
\varepsilon_{\rm finite}= \sum m_S^4\ln m_s +3 \sum m_V^4 \ln m_V - 2 \sum m_F^4\ln m_F.
\label{en-finite}
\end{equation}
Let us now calculate the vacuum pressure. This pressure is given by the formula \cite{zeld}
\begin{equation}
p=\frac{2\pi}{3}\int_0^{\infty}dk\frac{k^4}{\sqrt{k^2+m^2}}.
\label{pres-vac}
\end{equation}
On introducing the cutoff $\Lambda$ we have
\begin{eqnarray}
&&p=\frac{2\pi}{3}\int_0^{\Lambda}dk\frac{k^4}{\sqrt{k^2+m^2}}\nonumber \\
&&=\frac{2\pi}{3}m^4\left[\frac18\frac{\Lambda}{m}\left(\frac{2\Lambda^2}{m^2}\right)\sqrt{\frac{\Lambda^2}{m^2}+1}-\frac{\Lambda}{m}\sqrt{\frac{\Lambda^2}{m^2}+1}\right.\nonumber \\
&&\left.+\frac38\ln\left(\frac{\Lambda}{m}+\sqrt{\frac{\Lambda^2}{m^2}+1}\right)\right].
\label{pres-vac1}
\end{eqnarray}
On expanding this expression with respect to the small parameter $\frac{m}{\Lambda}$, we obtain
\begin{equation}
p=\frac{\pi}{6}\Lambda^4-\frac{\pi}{6}\Lambda^2m^2-\frac{7\pi}{48}m^4+\frac{\pi}{4}\ln 2+\frac{\pi}{4}m^4\ln\frac{\Lambda}{m}+o\left(\frac{m}{\Lambda}\right).
\label{pres-vac2}
\end{equation}
On then comparing the expressions (\ref{en2}) and (\ref{pres-vac2}), we see that the quartic divergence satisfies the  equation of state  
for  radiation $p=\frac13\varepsilon$, the quadratic divergence satisfies the equation of state $p=-\frac13\varepsilon$, which sometimes is identified with the so called string gas (see e.g. \cite{string,string1}), while the logarithmic divergence behaves as a cosmological constant with $p=-\varepsilon$. If all these divergences  cancel, then the finite part of the pressure is 
\begin{equation}
p_{\rm finite}=-(\sum m_s^4\ln m_s +3 \sum m_V^4 \ln m_V - 2 \sum m_F^4\ln m_F),
\label{p-fin}
\end{equation}
which also behaves as a cosmological constant.

\section{Running masses and anomalous mass dimensions}

In what follows we shall consider  models having only  particles with  spin zero and spin  $1/2$. 
If we include the interactions, the masses begin their running and the conservation of  the relations (\ref{quad}) and (\ref{log}) implies some
new restrictions on the masses and on the coupling constants. Namely, the conservation of the relation (\ref{quad}) gives
\begin{equation}
\sum\gamma_{m{\rm S}} = 2\sum\gamma_{m{\rm F}},
\label{quad-run}
\end{equation}
where $\gamma_m$ is the mass anomalous dimension defined as 
\begin{equation}
\gamma_m \equiv \mu\frac{\partial m^2}{\partial \mu},
\label{quad-run1}
\end{equation}
where as usual $\mu$ is the renormalization mass parameter. 
The conservation of the relation (\ref{log}) gives 
\begin{equation}
\sum m_S^2\gamma_{m{\rm S}} = 2\sum m_F^2\gamma_{m{\rm F}}.
\label{log-run}
\end{equation}
These relations coincide with those presented in paper \cite{Visser}.

We shall  here  derive the expressions for these anomalous mass dimensions. On considering 
our toy models with degenerate masses, we  shall not really use them explicitly. It will be enough to study shifts of masses induced by radiative corrections 
for different fields present in the models under consideration.  
However, when one considers models where particles with different masses are present, the formulas given in this section become necessary.

Our treatment of the anomalous mass dimensions in the presence of 
quadratic divergences is based on the approach presented in paper \cite{Jack}, which in turn uses the version of renormalization group formalism 
connected with  dimensional regularization \cite{Hooft}. 

Let us consider the model, including a Dirac spinor with a mass $M$ and a scalar field with a mass $m$.
\begin{equation}
L=\frac12\partial^{\mu}\phi\partial_{\mu}\phi-\frac{m^2\phi^2}{2}-\frac{\lambda\phi^4}{4!}+i\bar{\psi}\gamma^{\mu}\partial_{\mu}\psi-M\bar{\psi}\psi-g\bar{\psi}\psi\phi.
\label{Yukawa}
\end{equation}
The full propagator of the fermion field is given by 
\begin{equation}
S(p) = \frac{i}{\hat{p}-M-i\Sigma},
\label{ferm-prop}
\end{equation} 
where $\hat{A} \equiv \gamma_{\mu}A^{\mu}$ and 
where $\Sigma$ is the self-energy operator of the fermion field. 
This operator in the one-loop approximation is given by the formula
\begin{equation} 
\Sigma = g^2\int\frac{d^dk}{(2\pi)^d}\frac{1}{(p-k)^2-m^2}\frac{\hat{k}+M}{k^2-M^2} = \hat{p}\Sigma_1+M\Sigma_2.
\label{self-en}
\end{equation}
Here $d$ is the dimensionality of the spacetime such that 
\begin{equation}
d= 4-\varepsilon.
\label{dimen}
\end{equation}
Let us first calculate  the term $\Sigma_2$:
\begin{equation}
\Sigma_2 = g^2\int \frac{d^dk}{(2\pi)^{d}}\frac{1}{((p-k)^2-m^2)(k^2-M^2)}.
\label{sigma2}
\end{equation}
On making a Wick rotation, we obtain the integral on the Euclidean momenta:
\begin{equation}
\Sigma_2 = ig^2\int \frac{d^dk_E}{(2\pi)^{d}}\frac{1}{((p-k)_E^2+m^2)(k_E^2+M^2)}.
\label{sigma21}
\end{equation}
 We are interested only in the divergent part of this integral. Thus, we can neglect the masses in the denominator.
 On using the formula
 \begin{equation}
 \frac{1}{a} = \int_0^{\infty}e^{-\alpha a}d\alpha,
 \label{alpha}
 \end{equation}
 the Gaussian integration, the formula connecting the Euler $B$ and $\Gamma$ functions
 \begin{equation}
 B(a,b) \equiv \int_0^1 dx x^{a-1}(1-x)^{b-1} = \frac{\Gamma(a)\Gamma(b)}{\Gamma(a+b)},
 \label{Euler}
 \end{equation}
 and the fact that 
 \begin{equation}
 \Gamma(\varepsilon) = \frac{1}{\varepsilon}+\cdots,
 \label{Gamma}
 \end{equation}
 we arrive to the expression 
 \begin{equation}
 \Sigma_2 = \frac{ig^2}{8\pi^2\varepsilon}.
 \label{sigma22}
 \end{equation}
To find  $\Sigma_1$, we shall take the $\frac14{\rm Tr} (\hat{p} \Sigma)$. Then
 \begin{equation}
 \Sigma_1 = \frac{g^2}{p^2}\int\frac{d^dk}{(2\pi)^{d}}\frac{kp}{((p-k)^2-m^2)(k^2-M^2)}.  
 \label{sigma11}
 \end{equation}
 Using the identity
 \begin{equation}
 pk = \frac12(p^2+k^2-(p-k)^2),
 \label{scal-prod}
 \end{equation}
we transform the expression (\ref{sigma11}) as 
\begin{eqnarray}
&&\Sigma_1=\frac{g}{2p^2}\int\frac{d^dk}{(2\pi)^{d}}\left(\frac{1}{k^2-M^2}-\frac{1}{(p-k)^2-m^2}\right.\nonumber \\
&&\left.-\frac{m^2-M^2-p^2}{((p-k)^2-m^2)(k^2-M^2)}\right).
\label{sigma12}
\end{eqnarray}
This expression only contains  a logarithmic divergence. 
On making a Wick rotation, integrating in the Euclidean momentum space and keeping only the poles in $\varepsilon$, we obtain
\begin{equation}
\Sigma_1 = \frac{ig^2}{16\pi^2\varepsilon}.
\label{sigma13}
\end{equation}
Thus, 
\begin{equation}
\Sigma = \hat{p}\frac{ig^2}{16\pi^2\varepsilon}+ M\frac{ig^2}{8\pi^2\varepsilon}.
\label{sigma0}
\end{equation}

On substituting the formula (\ref{sigma0}) into Eq. (\ref{ferm-prop}) we see that
the fermion propagator in the one-loop approximation is 
\begin{equation}
S(p) = \frac{i}{\hat{p}\left(1+\frac{g^2}{16\pi^2\varepsilon}\right)-M\left(1-\frac{g^2}{8\pi\varepsilon}\right)}.
\label{ferm-prop1}
\end{equation}
In the same approximation, this propagator can be rewritten as 
\begin{equation}
S(p)=\frac{i\left(1-\frac{g^2}{16\pi^2\varepsilon}\right)}{\hat{p}-M\left(1-\frac{3g^2}{16\pi^2\varepsilon}\right)}.
\label{ferm-prop2}
\end{equation}
Thus, we see that the shift of the mass $M$ is 
\begin{equation}
\delta M = -\frac{3g^2 M}{16\pi^2\varepsilon}.
\label{shift-mass}
\end{equation}

To compensate this shift, we should introduce a counter-term into the Lagrangian, or in other terms, we should introduce 
a bare mass $M_B$ which is connected with the renormalized mass $M$ through the relation  
\begin{equation}
M_B = Z_M M,
\label{bare-mass}
\end{equation} 
where 
\begin{equation}
Z_M = 1+\frac{3g^2}{16\pi^2\varepsilon}.
\label{Z-M}
\end{equation}

Further, to have a canonically normalized fermion field, or in other words, to compensate a non-trivial divergent factor in the numerator of the formula (\ref{ferm-prop2}), we should introduce a bare fermion field
\begin{equation}
\psi_B = Z_{\psi}^{1/2}\psi,
\label{fermion-bare}
\end{equation}
where 
\begin{equation}
Z_B = 1+\frac{g^2}{16\pi^2\varepsilon}.
\label{fermion-bare1}
\end{equation}

On now, following the scheme elaborated in paper \cite{Hooft}, we introduce and calculate the anomalous mass dimension for the fermion mass $M$.
Let us remember that when we use the dimensional regularization, the renormalized quantities depend on the renormalization mass parameter $\mu$. At the same time 
the bare quantities depend on the regularization parameter $\varepsilon$, but do not depend on the renormalization mass parameter $\mu$. 
Thus, we can write down a general equation
\begin{equation}
\mu\frac{\partial}{\partial \mu} M_B = \mu\left(\frac{\partial}{\partial\mu}Z_M\right) M+Z_M\mu\frac{\partial}{\partial\mu}M = 0.
\label{renormgroup}
\end{equation}
Generally, the renormalization constant $Z_M$ has the following structure:
\begin{equation}
Z_M = 1+\sum_{n=1}^{\infty}\frac{a_n}{\varepsilon^n}.
\label{Z-M1}
\end{equation}
On introducing 
\begin{equation}
\gamma_M \equiv \mu\frac{\partial M}{\partial\mu},
\label{gamma-M}
\end{equation}
we can rewrite Eq. (\ref{renormgroup}) as follows:
\begin{equation} 
\gamma_M \left(1+\sum_{n=1}^{\infty}\frac{a_n}{\varepsilon^n}\right) + M \sum_{n=1}^{\infty}\mu\frac{\partial a_n}{\partial\mu}\frac{1}{\varepsilon^n}=0.
\label{Z-M2}
\end{equation}
For the case wherein the residues $a_n$ depend only on the Yukawa coupling constant $g$, Eq. (\ref{Z-M2}) becomes
\begin{equation}
\gamma_M \left(1+\sum_{n=1}^{\infty}\frac{a_n}{\varepsilon^n}\right) + M \sum_{n=1}^{\infty}\mu\frac{\partial g^2}{\partial\mu} \frac{da_n}{dg^2}\frac{1}{\varepsilon^n}=0.
\label{Z-M3}
\end{equation}
We now introduce the $\beta$ - function for the Yukawa constant $g$:
\begin{equation}
\beta_g \equiv \mu\frac{\partial g^2}{\partial\mu} + \varepsilon g^2.
\label{beta-g}
\end{equation}
Then Eq. (\ref{Z-M3}) reads:
\begin{equation}
\gamma_M \left(1+\sum_{n=1}^{\infty}\frac{a_n}{\varepsilon^n}\right) + M (\beta_g-\varepsilon g^2)\sum_{n=1}^{\infty} \frac{da_n}{dg^2}\frac{1}{\varepsilon^n}=0.
\label{Z-M4}
\end{equation}
The above equation should be correct in any order in $\varepsilon$. To the zeroth order it gives:
\begin{equation}
\gamma_M = g^2\frac{d a_1}{dg^2}.
\label{gamma-M1}
\end{equation} 
From Eq. (\ref{Z-M}) we immediately obtain 
\begin{equation}
\gamma_M = \frac{3g^2 M}{16\pi^2}.
\label{gamma-M2}
\end{equation}

The calculation of the analogous quantity for the scalar field is more complicated because the mass renormalization in this case includes quadratic divergences.
To treat them, we shall follow the approach developed in paper \cite{Jack}.  
The full propagator of the scalar field is 
\begin{equation}
D(p) = \frac{i}{p^2-m^2-i\Pi},
\label{scal-prop}
\end{equation}
where $\Pi$ is the self-energy operator. The contribution of the scalar field self-interaction to the one-loop order in the operator $\Pi$ is 
\begin{equation}
\Pi = \frac{\lambda\mu^{\varepsilon}}{2}\int\frac{d^dk}{(2\pi)^d}\frac{1}{k^2-m^2}.
\label{self-scal}
\end{equation} 
Let us note  that we here include  the factor $\mu^{\varepsilon}$ to provide the correct dimensionality of $\Pi$. We did not include such a factor on calculating the self-energy of the fermion, because there only logarithmic divergences were present and this factor disappeared in the limit $d \rightarrow 4$. Here, in the presence of quadratic divergences the factor $\mu^{\varepsilon}$ becomes crucial. A direct calculation gives 
\begin{equation}
\Pi = \frac{-i\lambda\mu^{\varepsilon}m^{d-2}}{2(4\pi)^{\frac{d}{2}}}
\Gamma\left(1-\frac{d}{2}\right).
\label{self-scal1}
\end{equation} 
One can see that this expression has the pole at $d=4$ and also the pole at $d=2$, corresponding to quadratic divergence \cite{Jack}. Indeed, it is well known that in the theory with the Lagrangian (\ref{Yukawa}) the index of divergence of a diagram $G$, $\omega(G)$ is 
\begin{equation}
\omega(G)=4-\frac32E_F-E_B,
\label{index}
\end{equation}
where $E_F$ is a number of the external fermion lines and $E_B$ is a number of the external boson lines. Thus, the diagrams  with $E_F=0, E_B =2$ are quadratically divergent. Let us now  consider $d$-dimensional spacetime. In this case, the formula (\ref{index}) is replaced by 
\begin{equation}
\omega(G) = (d-4)L+4-\frac32E_F-E_B,
\label{index1}
\end{equation}
where $L$ is the number of loops. Let us again consider  a diagram with $E_F=0, E_B=2$. This diagram, which is quadratically divergent at $d=4$ becomes logarithmically divergent ($\omega(G)=0$) at $d = 4-\frac{2}{L}$. That means that the quadratic divergence is represented as a pole of the quantity
\begin{equation}
\varepsilon^{(L)} = 4-d-\frac{2}{L},
\label{index2}
\end{equation}
and in the case of the one-loop approximation 
\begin{equation}
\varepsilon^{(1)} = 2-d.
\label{index3}
\end{equation}
Thus, expanding the expression (\ref{self-scal1}) around $d=4$, we have 
\begin{equation}
i\Pi_{d\rightarrow 4} = -\frac{\lambda m^2}{16\pi^2\varepsilon},
\label{self-scal2}
\end{equation}
while expansion of the same expression at $d \rightarrow 2$ gives 
\begin{equation}
i\Pi_{d\rightarrow 2} = \frac{\lambda\mu^2}{4\pi\varepsilon^{(1)}}.
\label{self-scal3}
\end{equation}
Thus, the infinite shift of the mass squared in the full propagator of the scalar field due to its self-interaction is 
\begin{equation}
\delta m^2 = -m^2\frac{\lambda}{16\pi^2\varepsilon}+\mu^2\frac{\lambda}{4\pi\varepsilon^{(1)}}.
\label{self-scal3}
\end{equation} 

We can analogously calculate  the contribution of the fermion loop to the self-energy of the scalar field.
\begin{equation}
\Pi = -g^2\mu^{\varepsilon}\int\frac{d^dk}{(2\pi)^d}\frac{{\rm Tr}[(\hat{k}+\hat{p}+M)(\hat{k}+M)]}{[(k+p)^2-M^2][k^2-M^2]}.
\label{ferm-loop}
\end{equation}
Calculation of this integral gives
\begin{eqnarray}
&&-i\Pi=\frac{4g^2M^{d-2}\mu^{\varepsilon}}{(4\pi)^{\frac{d}{2}}}\Gamma\left(1-\frac{d}{2}\right)\nonumber \\
&& +\frac{2g^2\mu{\varepsilon}(p^2-4M^2)}{(4\pi)^{\frac{d}{2}}}\Gamma\left(\frac{\varepsilon}{2}\right)\frac{\left[\Gamma\left(\frac{d}{2}-1\right)\right]^2}{\Gamma(d-2)}.
\label{ferm-loop1}
\end{eqnarray}
The logarithmic divergence is now
\begin{equation}
-i\Pi_{d\rightarrow 4} = \frac{g^2p^2}{4\pi^2\varepsilon}-\frac{3g^2M^2}{2\pi^2\varepsilon},
\label{ferm-loop2}
\end{equation}
while the quadratic divergence is 
\begin{equation}
-i\Pi_{d\rightarrow2} = \frac{2g^2\mu^2}{\pi\varepsilon^{(1)}}.
\label{ferm-loop3}
\end{equation}
The scalar field propagator corrected by the fermion loop is 
\begin{eqnarray}
&&D(p) = \frac{i}{p^2-m^2+\frac{g^2p^2}{4\pi^2\varepsilon}-\frac{3g^2M^2}{2\pi^2\varepsilon}+\frac{2g^2\mu^2}{\pi\varepsilon^{(1)}}}\nonumber \\
&&=\frac{i\left(1-\frac{g^2}{4\pi^2\varepsilon}\right)}{p^2-\left(m^2+\frac{3g^2M^2}{2\pi^2\varepsilon}-\frac{g^2m^2}{4\pi^2\varepsilon}-\frac{2g^2\mu^2}{\pi\varepsilon^{(1)}}\right)}.
\label{ferm-loop4}
\end{eqnarray}
Thus, the mass squared of the scalar field is shifted as
\begin{equation}
\delta m^2 = \frac{3g^2M^2}{2\pi^2\varepsilon}-\frac{g^2m^2}{4\pi^2\varepsilon}-\frac{2g^2\mu^2}{\pi\varepsilon^{(1)}}.
\label{ferm-loop5}
\end{equation}
On combining the last equation with Eq. (\ref{self-scal3}), we obtain the full mass shift:
\begin{equation}
\delta m^2 = -m^2\frac{\lambda}{16\pi^2\varepsilon}+\mu^2\frac{\lambda}{4\pi\varepsilon^{(1)}}+\frac{3g^2M^2}{2\pi^2\varepsilon}-\frac{g^2m^2}{4\pi^2\varepsilon}-\frac{2g^2\mu^2}{\pi\varepsilon^{(1)}}.
\label{self-scal-full}
\end{equation} 
To compensate this shift, we introduce a bare scalar field mass following the procedure elaborated in the paper \cite{Jack}:
\begin{equation}
m_B^2 = Z_m m^2+Z_{\mu}\mu^2,
\label{bare-scal}
\end{equation}
where the renormalization constants in the one-loop approximation are 
\begin{equation}
Z_m = 1+\frac{\lambda}{16\pi^2\varepsilon}-\frac{3g^2M^2}{2\pi^2m^2\varepsilon}+\frac{g^2}{4\pi^2\varepsilon}
\label{Z-m}
\end{equation}
and 
\begin{equation}
Z_{\mu} = -\frac{\lambda}{4\pi\varepsilon^{(1)}}+\frac{2g^2}{\pi\varepsilon^{(1)}}.
\label{Z-mu}
\end{equation}

Further, to have a canonical normalization of the scalar field, we introduce a bare field as follows:
\begin{equation}
\phi_B = Z_{\phi}^{1/2}\phi,
\label{scal-bare}
\end{equation}
where 
\begin{equation}
Z_{\phi} = 1+\frac{g^2}{4\pi^2\varepsilon}.
\label{scal-bare1}
\end{equation}

On now introducing the anomalous mass dimension 
\begin{equation}
\gamma_m \equiv \mu\frac{\partial m^2}{\partial \mu},
\label{gamma-m}
\end{equation}
and requiring the independence of the bare mass (\ref{bare-scal}) on the renormalization mass parameter $\mu$ and using the explicit expressions (\ref{Z-m}) and (\ref{Z-mu}), we obtain  in the one-loop approximation 
\begin{equation}
\gamma_m = \frac{\lambda m^2}{16\pi^2}-\frac{3g^2M^2}{2\pi^2}+\frac{g^2m^2}{4\pi^2}-\frac{\lambda\mu^2}{4\pi}+\frac{2g^2\mu^2}{\pi}.
\label{gamma-m1}
\end{equation}

Let us also include some pseudoscalar fields into our model. The interaction between a scalar and the fermion is described by the following term in the Lagrangian:
\begin{equation}
L = -h\bar{\psi}\gamma^5\psi\chi.
\label{pseudo}
\end{equation}
Here 
\begin{equation}
(\gamma^5)^2 = -1,\ \gamma^5 \hat{k} \gamma^5 = \hat{k}.
\label{gamma5}
\end{equation}
On repeating the preceding calculations  and using the formulae (\ref{gamma5}), we see that the contribution of the interaction (\ref{pseudo}) to the anomalous mass dimension of the fermion field is 
\begin{equation}
\gamma_M = \frac{h^2M}{16\pi^2}.
\label{pseudo1}
\end{equation}
The contribution of the fermion loop to the anomalous mass dimension of the pseudoscalar field $\chi$ is 
\begin{equation}
\gamma_{m_{\chi}}=-\frac{h^2M^2}{2\pi^2}+\frac{h^2m^2}{4\pi^2}+\frac{2h^2\mu^2}{\pi}.
\label{pseudo2-cor}
\end{equation}

\section{Contribution of potential terms into the vacuum energy and the auxiliary fields}

When we switch on the interactions and require the cancellation of ultraviolet divergences in the expression for vacuum energy, we should consider not only the mass sum rules, but also the potential terms. The contributions of the potential terms 
to the vacuum energy density have the following structure
\begin{equation}
E_{\rm pot}=\frac{\langle 0| T (V \exp(i\int d^4x L_{\rm int}))|0\rangle}{\langle 0| T  \exp(i\int d^4x L_{\rm int})|0\rangle}.
\label{poten-vac}
\end{equation}   
Here, $L_{\rm int}$ is the interaction Lagrangian   and the exponent should be expanded up to necessary order in the perturbation theory while $V$ represents  potential terms. 
For the term 
\begin{equation}
V = \lambda\phi^4,
\label{poten-vac1}
\end{equation}
to obtain the result in the two-loop approximation which we study in the present paper, it is enough to take only the zeroth order of the expansion of the exponent of the action in the formula (\ref{poten-vac}). 
The corresponding contribution is equal to
\begin{equation}
E_1 = -3\lambda I^2, 
\label{pot-quart}
\end{equation}
where the integral $I$ is  defined as 
\begin{equation}
I = \int\frac{dk}{k^2-m^2}.
\label{integral}
\end{equation} 

The contribution of the Yukawa interaction term  is given by the structure
\begin{equation}
E_2 = \langle 0| T (g\bar{\psi}\psi\phi\times (-ig)\bar{\psi}\psi\phi|0\rangle,
\label{pot-Yuk}
\end{equation}
where the second factor $(-ig)\bar{\psi}\psi\phi$ comes from the first-order term in the expansion of the $T$-exponent. This contribution (for the case of a Majorana spinor) is 
equal to 
\begin{equation}
E_2=2g^2\int\frac{{\rm Tr}[(\hat{p}+\hat{k}+M)(\hat{k}+M]}{[(p+k)^2-M^2][k^2-M^2][p^2-m^2]},
\label{pot-Yuk1}
\end{equation}
where $M$ is the fermion mass and $m$ is the scalar mass. A simple calculation shows that for the case of the Wess-Zumino model, when $m=M$ and there are 
well-known relations between the coupling constants \cite{WZ}, the quartic divergences present in the contributions (\ref{pot-quart}) and (\ref{pot-Yuk1}) do not cancel each other (we shall present detailed calculations in the next section). Namely, the contribution of the spinors is twice  that of the scalars. The reason for this mismatch was already discussed in the Introduction. The point is that the number of fermion degrees of freedom is doubled off shell. To compensate this effect, we should introduce the auxiliary scalar fields as  is done in  supersymmetric models. A simple example shows that this exactly gives the doubling of the leading contribution to vacuum energy. Indeed, let us consider a model with the Lagrangian
\begin{equation}
L = \frac12(\partial_{\mu}\phi)^2+\frac{F^2}{2}+hF\phi^2.
\label{toy-aux}
\end{equation}
On shell this theory is equivalent to the theory where the auxiliary field is excluded $F$ by means to the equation of motion
\begin{equation}
F+h\phi^2=0
\label{on-shell}
\end{equation}
and which has a Lagrangian
\begin{equation}
L = \frac12(\partial_{\mu}\phi)^2-\frac12h^2\phi^4.
\label{toy-aux1}
\end{equation}
It follows from Eq. (\ref{pot-quart}) that vacuum energy in the theory with the Lagrangian (\ref{toy-aux1}) is equal to 
\begin{equation}
E_{\rm vacuum} = -\frac32h^2I^2.
\label{toy-aux2}
\end{equation}

Let us calculate an analogous (two-loop) contribution to vacuum energy in the model with the Lagrangian  (\ref{toy-aux}).
It is equal to 
\begin{equation}
E_{\rm vacuum}=\langle 0|T(-hF\phi^2\times (ih)F\phi^2|0\rangle = -3h^2I^2.   
\label{toy-aux3}
\end{equation}
Here we have used the fact that the propagator of the auxiliary field in the massless theory is given \cite{Iliop-Zum} by
\begin{equation}
\langle 0|T(FF)|0\rangle = i.
\label{prop-F}
\end{equation}

We see that in this case the result is doubled because of the effective doubling of the number of degrees of freedom.
One can check also that the contributions to the self-energy operator of the scalar field to the order of $h^2$ coincide in the models 
with the Lagrangians (\ref{toy-aux}) and (\ref{toy-aux1}):
\begin{eqnarray}
&&\Big\langle 0\Big|T\left(\phi\phi\times\left(-\frac{ih^2\phi^4}{2}\right)\right)\Big|0\Big\rangle\nonumber\\
&&  = \Big\langle 0\Big|T\left(\phi\phi\times\frac12(ihF\phi^2)^2\right)\Big|0\Big\rangle=6h^2I.
\label{self-coincide}
\end{eqnarray}
Thus, the requirement of the explicit account of auxiliary fields arises only in the diagrams possessing quartic ultraviolet divergences and including only boson propagators, because their contribution is proportional to the number of degrees of freedom present off shell in the model under consideration. This fact gives us a practical recipe: when one calculates vacuum energy contribution of the scalar field diagrams, having the shape of ``eight'', one should multiply it by the factor $2$. 

Concluding this section, we wish to make one more comment. In
the action (\ref{toy-aux}) the term $\frac{F^2}{2}$ is also present. One can consider this term as a part of the kinetic energy. The contribution of this term into vacuum energy is given by the formula 
\begin{eqnarray}
&&\Big\langle 0\Big|T\left(-\frac{F^2}{2}\times (ih)F\phi^2|0\times (ih)F\phi^2|0\times\frac12\right)\Big|0\Big\rangle\nonumber \\
&&=+\frac32h^2I^2.
\label{correction}
\end{eqnarray}
Thus, on summing (\ref{correction}) and (\ref{toy-aux3}) we reproduce the result (\ref{toy-aux2}). This means that, in the end, the results for vacuum energy 
in the model (\ref{toy-aux}) with an auxiliary field and in the model (\ref{toy-aux1}), where the auxiliary field is eliminated, coincide.
However, the expressions for the contributions of the potential energy and of the kinetic energy do not coincide separately. As we have already mentioned in the 
Introduction, a similar effect can be observed in the supersymmetric Wess-Zumino model. If we consider the formalism in the absence of auxiliary fields, vacuum energy is still equal to zero, but the potential and kinetic energy are not equal to zero separately.

\section{A model with one Majorana and two scalar fields}
Let us consider a model with a Majorana field $\psi$ and two scalar fields $A$ and $B$.
All the fields have the same mass $m$ and the interaction is given by 
\begin{eqnarray}
&&H_{\rm int} = \lambda_{1}A^4+\lambda_2B^4+\lambda_3A^2B^2\nonumber \\
&&+g_1\bar{\psi}\psi A+g_2\bar{\psi}\psi B\nonumber \\ 
&&+mh_1A^3+mh_2B^3+mh_3A^2B+mh_4AB^2.
\label{inter}
\end{eqnarray}
The two tadpole diagrams for fields $A$ and $B$ should be cancelled to avoid the necessity of introducing linear in fields terms into the Lagrangian.
The tadpole for the field $A$ arises due to the contraction of this field with the vertices $A^3, AB^2$ and $\bar{\psi}\psi A$. All these contributions are proportional 
to the integral (\ref{integral}). 
The corresponding combinatorial factors are $3mh_1$, for $A^3$, $mh^4$ for $AB^2$ and $-4mg_1$ for the vertex $\bar{\psi}\psi A$. The last contribution arises due to the trace of the fermion propagator which is proportional to the mass $m$.
Thus, the cancellation of the tadpole diagram for $A$ requires
\begin{equation}
3h_1+h_4=4g_1.
\label{tadpole}
\end{equation}
Similarly the vanishing of the tadpole for the field $B$ requires
\begin{equation}
3h_2+h_3=4g_2.
\label{tadpole1}
\end{equation}

Now the self-energy operator for the propagator of the field $A$ obtains the contributions from the vertex $A^4$, from the vertex $A^2B^2$ and from the pair of vertexes 
$\bar{\psi}\psi A$, $A^3$, $A^2B$ and $AB^2$. The contributions of two quartic vertexes are both proportional to the integral $I$. The corresponding coefficients are 
$12\lambda_1$ and $2\lambda_3$. 
The contribution of the fermion loop is 
\begin{equation}
C_1=-2g_1^2{\rm Tr} \int\frac{{\rm Tr} ((\hat{p}+\hat{k}+m)(\hat{k}+m)}{[(p+k)^2-m^2][k^2-m^2]}dk, 
\label{fermion}
\end{equation}
where the factor $2$ arises due to the Majorana nature of the fermion.
Then 
\begin{eqnarray}
&&C_1=-8g_1^2\int\frac{k^2+kp+m^2}{[(p+k)^2-m^2][k^2-m^2]}dk\nonumber \\
&&=-4g_1^2\int\frac{(k^2-m^2)+((k+p)^2-m^2)-p^2+4m^2}{[(p+k)^2-m^2][k^2-m^2]}dk\nonumber \\
&&=-8g_1^2 I +(4p^2-16m^2)g_1^2K,
\label{fermion1}
\end{eqnarray} 
where
\begin{equation}
K =\int\frac{dk}{[(p+k)^2-m^2][k^2-m^2]}.
\label{integral1}
\end{equation} 
Quadratic divergences present in the integral $I$ should be canceled because such divergences do not arise in the 
self-energy correction to the fermion propagator. Thus, we have 
\begin{equation}
12\lambda_1+2\lambda_3 - 8g_1^2=0
\label{quadratic}
\end{equation}
and, analogously,
\begin{equation}
12\lambda_2+2\lambda_3 - 8g_2^2=0.
\label{quadratic1}
\end{equation}
Now the contribution of the pairs of the triple scalar vertices is 
\begin{equation}
C_2 = (18h_1^2+4h_3^2+2h_4^2)m^2K.
\label{scalar}
\end{equation}
Thus, the propagator of the scalar field $A$ in the one-loop approximation has the form
%\begin{onecolumn}
\begin{eqnarray}
&&(G_A)^{-1}=-i(p^2(1-4ig_1^2K)\nonumber \\
&&-m^2(1+i(-16g_1^2+18h_1^2+4h_3^2+2h_4^2)K).
\label{propag}
\end{eqnarray}
%\end{onecolumn}
Normalizing as usual the wave function, i.e. making the coefficient at $p^2$ equal to $1$, we obtain
\begin{equation}
G_A= \frac{i}{p^2-m^2(1+i(-12g_1^2+18h_1^2+4h_3^2+2h_4^2)K)}.
\label{propag1}
\end{equation}
Thus, this effective shift of the mass squared given by 
\begin{equation}
im^2(-12g_1^2+18h_1^2+4h_3^2+2h_4^2)K
\label{shift}
\end{equation}
defines the running of the mass for the scalar field $A$.
The analogous shift for the second scalar field is 
\begin{equation}
im^2(-12g_2^2+18h_2^2+4h_4^2+2h_3^2)K.
\label{shift1}
\end{equation}
The self-energy contribution to the fermion propagator is 
\begin{eqnarray}  
&&C_3=4(g_1^2+g_2^2)\int\frac{\hat{k}+m}{[(p-k)^2-m^2][k^2-m^2]}\nonumber \\
&&=(g_1^2+g_2^2)(2\hat{p}+4m)K,
\label{fermion2}
\end{eqnarray}
where the factor $4$ arises due to the Majorana nature of the fermion. 
The fermion propagator is now  
\begin{equation}
G_F = \frac{i}{\hat{p}(1-2i(g_1^2+g_2^2)K)-m(1+4i(g_1^2+g_2^2)K}.
\label{prop-ferm}
\end{equation}
On normalizing the term at $\hat{p}$, we obtain
\begin{equation}
G_F=\frac{i}{\hat{p}-m(1+6i(g_1^2+g_2^2)K}.
\label{prop-ferm1}
\end{equation}
The shift of the mass squared is 
\begin{equation}
12im^2(g_1^2+g_2^2)K.
\label{shift2}
\end{equation}
The running of the masses and, hence, the shifts (\ref{shift}), (\ref{shift1}) and (\ref{shift2}) should be equal and we obtain two equations:
\begin{equation}
18h_1^2+4h_3^2+2h_4^2-12g_1^2 = 12(g_1^2+g_2^2)
\label{running}
\end{equation}
and 
\begin{equation}
18h_2^2+4h_4^2+2h_3^2-12g_2^2 = 12(g_1^2+g_2^2).
\label{running1}
\end{equation}

Let us now consider the contribution of the potential term (\ref{inter}) to vacuum energy. 
The contribution of the quartic terms is 
\begin{equation}
E_1 = (3\lambda_1 + 3\lambda_2 + \lambda_3)I^2.
\label{pot-en}
\end{equation}
The contribution coming from the two scalar-fermion vertices is given by the integral 
\begin{eqnarray}
&&E_2=-8\int\frac{{\rm Tr}((\hat{p}+\hat{k}+m)(\hat{k}+m)}{[(p+k)^2-m^2][k^2-m^2][p^2-m^2]}dkdp\nonumber \\
&&=-4(g_1^2+g_2^2)I^2-12m^2(g_1^2+g_2^2)L,
\label{pot-en1}
\end{eqnarray}
where 
\begin{equation}
L = \int \frac{dk dp}{[(p+k)^2-m^2][k^2-m^2][p^2-m^2]}.
\label{L}
\end{equation}
The contribution to vacuum energy of the triple scalar interactions is 
\begin{equation}
E_3=m^2(6h_1^2+6h_2^2+2h_3^2+2h_4^2)L.
\label{pot-en2}
\end{equation}
We can now observe   that the sum 
\begin{equation}
2E_1+E_2+E_3 = 0,
\label{vacuum-canc}
\end{equation}
provided Eqs. (\ref{quadratic}), (\ref{quadratic1}), (\ref{running}) and (\ref{running1}) are satisfied. The coefficient $2$ in front of the term $E_1$ is 
 introduced to take into account the fact that the number of  boson and fermion degrees of freedom should be equal also off shell.  It is equivalent 
 to the introduction of two auxiliary fields in supersymmetric models, as was explained in the preceding section. 

On now substituting the expressions for $g_1$ and $g_2$ from Eqs. (\ref{tadpole}) and (\ref{tadpole1}) into Eqs. (\ref{running}) and (\ref{running1}), we obtain the following pair 
of the consistency conditions on the constants $h_1,h_2,h_3$ and $h_4$:
\begin{equation}
18h_1^2-27h_2^2+13h_3^2+2h_4^2-36h_1h_4-18h_2h_3=0,
\label{consist}
\end{equation}
\begin{equation}
18h_2^2-27h_1^2+13h_4^2+2h_3^2-36h_2h_3-18h_1h_4=0.
\label{consist1}
\end{equation}

These equations are homogeneous in $h_1,h_2,h_3$ and $h_4$. Thus, we can fix $h_1 = 1$ and  we can then change the value of $h_2$. Then we shall have a system of two quadratic equations for $h_3$ and $h_4$. This system, which is equivalent to one quartic equation for one variable, is solvable analytically but the solutions are very cumbersome. Thus, we  shall just  present some numerical solutions. In any case we have four solutions, two of them are complex and two of them are real. 
We shall only take real solutions into account.\\
\begin{eqnarray*}
&&h_2 = 1,\\
&&h_3=h_4 \approx 3.8\\
&&{\rm or}\\
&&h_3=h_4\approx -0.2. 
\end{eqnarray*}
Then
\begin{eqnarray*}
&&h_2=0.9\\
&&h_3\approx 3.56,\, \, h_4\approx 3.58\\
&&{\rm or}\\
&&h_3\approx -0.46,\, \, h_4\approx 0.17.
\end{eqnarray*} 
Then
\begin{eqnarray*}
&&h_2=10/9\\
&&h_3\approx 3.98,\ \ h_4\approx 3.96\\
&&{\rm or}\\
&&h_3\approx0.2,\ \ h_4\approx-0.5.
\end{eqnarray*}
\begin{eqnarray*}
&&h_2=1/2\\
&&h_3\approx 2.8,\ \ h_4\approx 2.9\\
&&{\rm or}\\
&&h_3\approx-1.2,\ \ h_4\approx 1.2.
\end{eqnarray*}
\begin{eqnarray*}
&&h_2=2\\
&&h_3\approx 5.8,\ \ h_4\approx 5.6\\
&&{\rm or}\\
&&h_3\approx2.5,\ \ h_4\approx-2.4.
\end{eqnarray*}
\begin{eqnarray*}
&&h_2=1/10\\
&&h_3\approx 2.09,\ \ h_4\approx 2.26\\
&&{\rm or}\\
&&h_3\approx-1.8,\ \ h_4\approx 1.9.
\end{eqnarray*}
\begin{eqnarray*}
&&h_2=10\\
&&h_3\approx 22,\ \ h_4\approx 21\\
&&{\rm or}\\
&&h_3\approx 19,\ \ h_4\approx-18.
\end{eqnarray*}
Let us note that the negative values of the coupling constants are not essential because they are
in front of the odd (third) power of fields $A$ and $B$. The lower bound of the scalar field potential exists and 
is determined by the quartic terms with positive constants $\lambda_1,\lambda_2$ and $\lambda_3$. 

One can meanwhile introduce the quartic interactions using auxiliary fields in a manner similar to that used in the Wess-Zumino model. It is enough to introduce the following terms into 
 the Lagrangian instead of terms with quartic interactions:
\begin{equation}
 \frac{F^2}{2}+\frac{G^2}{2}+F(\sqrt{2\lambda_1}A^2-\sqrt{2\lambda_2}B^2)+G(\sqrt{2\lambda_3}+2\sqrt{\lambda_1\lambda_2})AB.
 \label{auxiliary}
 \end{equation}

 \section{Model with a Majorana field, a scalar field and a pseudoscalar field}
 
 Let us consider another toy model where the field $B$ is a pseudoscalar. In this case 
 $$
 h_2=h_3=0
$$
and the interaction between the pseudoscalar and the fermion is described by the Lagrangian
$$
g_2\bar{\psi}\gamma_5\psi B.
$$
In this case we have only one condition for the tadpole cancellation for the scalar field $A$ which coincides with that given by Eq. (\ref{tadpole}). 
The conditions for the cancellation of quadratic divergences in the propagators of the scalar and pseudoscalar fields are also the same 
(\ref{quadratic}) and (\ref{quadratic1}). However, the shifts of the mass squared for the fields $A,B$ and $\psi$ are different. They are proportional to
\begin{eqnarray}
&&\delta m^2_A \sim -12g_1^2+18h_1^2+2h_4^2,\nonumber \\
&&\delta m_B^2 \sim 4h_4^2+4g_2^2,\nonumber \\
&&\delta m^2_{\psi} \sim 12g_1^2-4g_2^2,
\label{pshifts}
\end{eqnarray}  
respectively.
Correspondingly, on requiring that the running of these three masses are the same, we obtain the following couple of equations:
\begin{equation}
-12g_1^2+18h_1^2+2h_4^2=4h_4^2+4g_2^2
\label{p-running}
\end{equation}
and 
\begin{equation}
4h_4^2+4g_2^2=
12g_1^2-4g_2^2.
\label{p-running1}
\end{equation}
From these two equations we obtain immediately
\begin{equation}
g_1=\pm h_1.
\label{p-running2}
\end{equation}
If 
$$
g_1=h_1
$$ 
then from Eq. (\ref{tadpole}) we obtain 
$$
h_4=g_1,
$$ 
which, in turn, implies 
$$g_1^2=g_2^2$$
and 
$$\lambda_1=\lambda_2.$$ 
If, instead,
\begin{equation}
g_1=-h_1,
\label{w-choice}
\end{equation}
then 
$$
h_4=-7g_1,
$$
which implies a negative value for $g_2^2$ or for $g_1^2$ as follows from the couple of equations  (\ref{p-running}) and (\ref{p-running1}).
Thus, the choice (\ref{w-choice}) should be discarded.

We have seen that for the case with one Majorana field, one scalar and one pseudoscalar we have less freedom in the choice of the coupling constants than in the case of 
two scalar fields and one Majorana field, but this choice is still broader than that in the Wess-Zumino model.

In principle, one can also consider  a model with one Majorana field and two pseudoscalar fields. In this case the triple scalar interactions do not exist and the constants $h_1,h_2,h_3$ and $h_4$ all are equal to zero. The requirement of the absence of quadratic divergences in the shifts of the mass squared of two pseudoscalar propagators 
is again given by Eqs. (\ref{quadratic}) and (\ref{quadratic1}). However, the shifts of the mass squared for this propagators are equal to zero, while the shift of the mass squared of the fermion propagator is given by 
$$
-4g_1^2-4g_2^2.
$$
Thus, the Yukawa interactions should vanish as well. Now, the conditions   (\ref{quadratic}) and (\ref{quadratic1}) can be satisfied if 
$$
\lambda_3 = -6\lambda_1=-6\lambda_2,
$$
but the corresponding quartic potential of these two pseudoscalar fields is unbounded from below and is hardly interesting.\\

\section{Models with non-degenerate masses}

It is interesting to find  toy models with  masses which are not degenerate. In this case it is necessary to consider at least four boson and four fermion degrees of freedom \cite{we}. 
The simplest models of this kind are those which include a certain number of ``triplets'' of the types described in two preceding sections, i.e. with degenerate masses inside  any triplet and with coupling constants (again, describing interactions within a triplet) which satisfy the relations obtained in the Sections V and VI. 
Naturally, in this case, if there are no interactions between the fields belonging  to different triplets, then  the Pauli-Zeldovich mechanism does work. If we introduce 
interactions  between different triplets with different masses, then the coupling constants should satisfy some constraints. 

We shall illustrate this by considering a model, where 
there are two triplets. Both triplets contain a Majorana fermion and two scalar fields. The mass of all the particles in the first triplet is equal to $m_1$, in the second triplet - $m_2$ and the coupling constants describing interactions within the triplets are chosen in such a way that vacuum energy is equal to zero. 
Let us then  introduce the following interaction Hamiltonian between fields belonging to  different triplets:
\begin{eqnarray}
&&H = \lambda_{AC}A^2C^2+\lambda_{AD}A^2D^2+\lambda_{BC}A^2C^2+\lambda_{BD}A^2D^2\nonumber \\
&&+g_{A\chi}\bar{\chi}\chi A+g_{B\chi}\bar{\chi}\chi B +g_{C\psi}\bar{\psi}\psi C+g_{D\psi}\bar{\psi}\psi D\nonumber \\
&&+h_{AC1}A^2C+h_{AC2}AC^2+h_{AD1}A^2D+h_{AD2}AD^2\nonumber \\
&&+h_{BC1}B^2C+h_{BC2}BC^2\nonumber \\
&&+h_{BD1}B^2D+h_{BD2}BD^2, 
\label{triplets}
\end{eqnarray} 
where the scalar fields $A$ and $B$ and the Majorana spinor $\psi$ belong to the first triplet, while the scalar fields $C$ and $D$ and the Majorana spinor 
$\chi$ belong to the second triplet. 

It is now possible  to find relations which constrain the choice of the coupling constants given in the interaction Hamiltonian (\ref{triplets}).  
However, this task is rather cumbersome and we shall postpone it and the construction of other models for future work \cite{future}.

\section{Concluding remarks}

In this paper we have studied the Pauli-Zeldovich mechanism for the cancellation of ultraviolet divergences in vacuum energy which is  associated with the fact 
that bosons and fermions produce contributions to it having opposite signs. In contrast with the preceding papers devoted to this topic where only  free fields were considered,  here we have taken interactions up to the lowest order of  perturbation theory into account. We have constructed a number simple toy models having
particles with spin $0$ and spin $1/2$, wherein masses of the particles are equal while  interactions can be quite non-trivial. 
To make calculations  simpler and more transparent, it was found useful to introduce some auxiliary fields. It appears that the presence of these fields is equivalent to the 
modification of some contributions of the physical fields to the vacuum expectation of the potential energy. We hope to construct more complicated models including particles with different masses and in the presence of  vector fields in  future work \cite{future}.

\section*{Acknowledgments}
We are grateful to A.~O.~Barvinsky, F.~Bastianelli and O.~V.~Teryaev for useful discussions.

\end{document}